\documentstyle[twocolumn,aps,prl]{revtex}
\include{epsf}
\epsfverbosetrue
\begin{document}
\twocolumn[\hsize\textwidth\columnwidth\hsize\csname@twocolumnfalse\endcsname

\draft

\title{Surface soft phonon and the $\sqrt{3}\times\sqrt{3}\leftrightarrow
3\times3$ phase transition in Sn/Ge(111) and Sn/Si(111).}
\author{Rub\'en P\'erez, Jos\'e Ortega and  Fernando Flores}
\address{Departamento de F\'{\i}sica Te\'orica de la Materia Condensada, 
Universidad Aut\'onoma de Madrid, E-28049 Madrid, Spain}
\date{manuscript LY7332, final version accepted in Physical Review Letters}
\maketitle

\begin{abstract}
Density Functional Theory (DFT) calculations show that 
the reversible Sn/Ge(111) $\sqrt{3}\times\sqrt{3} \leftrightarrow 3\times3$ 
phase transition can be described in terms of a surface soft phonon.
The isovalent Sn/Si(111) case does not display this transition  since 
the $\sqrt{3}\times\sqrt{3}$  phase is the stable
structure at low temperature, although it presents a partial softening
of the $3\times3$ surface phonon.
The rather flat energy surfaces for the atomic motion associated with 
this phonon mode in both cases explain the experimental similarities 
found at room temperature between these systems.
The driving force underlying the 
$\sqrt{3}\times\sqrt{3} \leftrightarrow 3\times3$
phase transition is shown to be associated with the 
electronic energy gain due to the Sn dangling bond rehybridization.
\end{abstract}


]

\narrowtext

A soft phonon is a vibrational mode of a crystalline material whose frequency
decreases as T falls, eventually reaching zero. At this point
the crystal is unstable in relation to the corresponding
atomic displacements and
undergoes a transition to a lower symmetry phase. Typical examples are the
high--low transition in quartz, the ferroelectric transition in
$BaTiO_3$ and the ferroelastic transition in $SrTiO_3$
\cite{putnis}.
On semiconductor surfaces, different mechanisms, like
Peierls transitions or Charge Density Waves, have been proposed as the origin of
two dimensional phase transitions. However, no soft phonon transition has been
unambiguously identified in these systems so far.

In this letter we consider 
the Sn/Ge(111) $\sqrt{3}\times\sqrt{3} \leftrightarrow 3\times3$ phase
transition, which is a prototypical example
of a T-induced reversible change of symmetry at a semiconductor surface.
The driving force underlying this phase transition has been under intense debate
since its discovery on the Pb/Ge(111) system \cite{nature}.              
Different physical mechanisms have been proposed to play an important role
in this transition: surface Fermi wavevector nesting or electron
correlations, leading to
the formation of a Charge Density Wave \cite{nature,carpinelli} at low T;
dynamical fluctuations
of an underlying $3\times3$ structure, giving rise to the
observation, on average, of a $\sqrt{3}\times\sqrt{3}$ symmetry
at room T \cite{avila};
and the interaction between the $3\times3$ periodicity and
Ge--substitutional defects that act as nucleation centers\cite{weitering}.
Recently, several groups \cite{uhrberg,ottaviano,yamanaka,themlin}
have also considered the isovalent Sn/Si(111) system and
have found strong similarities and some striking differences with the Ge case.
The electronic structure is very similar in both cases, with two major
components in the Sn $4d$ core level spectra
and two surface bands close to the Fermi level \cite{uhrberg},
associated with two
different Sn--dangling--bonds.
At variance with the Sn/Ge(111), where a transition to
a $3\times3$ pattern is clearly observed at low T, the Sn/Si(111)
surface shows a $\sqrt{3}\times\sqrt{3}$ pattern both in LEED and STM, 
for temperatures as low as 70 K \cite{uhrberg},
even in the presence of a significant density of Si--substitutional
defects \cite{ottaviano}.
Moreover, the two surface bands mentioned above are not as clearly resolved
at low T as they are in the Sn/Ge(111) case\cite{uhrberg}.
By analyzing theoretically  the Sn/Ge and Sn/Si systems, we
conclude that the $\sqrt{3}\times\sqrt{3} \rightarrow 3\times3$
transition is due to a surface soft phonon. 
The frequency of this mode in Sn/Ge(111) goes to zero for a
$\bar{k}$--vector corresponding to a reciprocal lattice vector of the
$3\times3$ periodicity. 
For Sn/Si(111) we find that the $\sqrt{3}\times\sqrt{3}$ is the stable structure
at low T, and therefore this system should not display the $\sqrt{3}\times\sqrt{3}
\rightarrow 3\times3$ transition.
The comparison of the Sn/Ge and Sn/Si systems with their corresponding 
H-covered cases shows that the driving force for this transition is the 
electronic energy gain associated
with the surface band splitting induced by the $3\times 3$ distortion.

The atomic displacements associated with the $\sqrt{3}\times\sqrt{3} \rightarrow
3\times3$ transition in Sn/Ge(111)
provide the clue for the soft mode responsible
for this transition.
The signature of the $3\times3$ structure is the upward displacement of
one of the three Sn-adatoms, and the corresponding downward movement of
the other two \cite{ortega}. Figure~1 and Table I show how these
 upward and downward movements are interconnected:
as one of the Sn atoms ($Sn_1$) moves upwards, the three Ge nearest
neighbours ($A_1$) follow this motion by moving towards it (with both upward and in--plane
displacements) . These Ge displacements force corresponding
in-plane movements of the Ge (labelled D$_2$) towards the 
position of $Sn_1$.
 Since the Ge--D$_2$ atoms are also bonded to 
first--layer Ge--atoms ($B_1$,$C_1$)
linked with the two
other Sn atoms ($Sn_2$ and $Sn_3$), these are forced to move downwards. 

Does the Sn/Si(111) surface present a similar $3\times3$ structure
at low T?.
We have studied this possibility by means of 
DFT calculations \cite{castep}. In this analysis, 
we start from a $\sqrt{3}\times\sqrt{3}$ structure (in
a $3\times3$ unit cell), where the 3 Sn atoms are equivalent, we select one of
them ($Sn_1$) and force it to move in the direction perpendicular to the
surface. For each of these displacements, the other Sn atoms ($Sn_2$ and $Sn_3$)
and all the semiconductor atoms are allowed to relax (up to the fifth
layer) to their zero
force positions under the constraint of the $Sn_1$ displacement.
These calculations were performed for both the Sn/Si and Sn/Ge systems
(see Figure~2).
The energy vs. $Sn_1$ displacement is very flat for Sn/Ge(111); the
ground state corresponds to a $3\times3$ geometry with a $Sn_1$
displacement of $\sim 0.18$ \AA, as found in
previous calculations \cite{ortega,gironcoli2000} and in agreement with the
experimental evidence \cite{sxrd1,sxrd2}.
The Sn/Si(111) surface, however, does not present
any stable $3\times3$ distortion, {\it i.e.}, the
$\sqrt{3}\times\sqrt{3}$ is the stable structure at low T.

The inset of Fig. 2 shows another interesting finding:
$Sn_2$ and $Sn_3$
move together in the
opposite direction to the constrained motion of the $Sn_1$ atom,
{\it for both the Sn/Ge(111) and Sn/Si(111) surfaces}. 
These results
suggest a correlated up/down motion of the Sn atoms \cite{avila} that, at a
sufficiently low temperature, is frozen into the $3\times3$
structure in the case of Ge (but not in the case of Si).
As reflected in Table I, the displacements associated with this motion   
are essentially localized in the
tetrahedra formed by a Sn atom and the three Ge of
the first layer bonded to it.
In order to analyze the corresponding normal modes 
associated with
the up/down motion of the tetrahedra,
we have considered a force constant model which includes effective interactions  
between first nearest neighbour tetrahedra:


\begin{equation}
F_i = -\alpha z_i + \beta \sum_{j,nn} (z_j - z_i)
\label{eqn:force}
\end{equation}
where $F_i$, the force acting on each tetrahedron,
is proportional to its center--of--mass displacement (with respect to the
$\sqrt{3}\times\sqrt{3}$ structure), $z_i$, and the differences in displacement
between nearest neighbours, $z_j - z_i$.

The force constants $\alpha$ and $\beta$ can be determined from 
the calculations presented above. These calculations provide
the accurate information
required for this purpose, since they are focused precisely on the deformations
we are analyzing here. 
Imposing that $F_2$ and $F_3$ are zero in the above
equations yields a relation between the force constants and the 
displacements $z_{1}$ and $z_{2}(=z_{3})$: $3\beta/ \alpha = - z_{2} /
(z_{2} - z_{1}) $.
For small displacements, ($-z_2$) is around
0.5$z_1$ for Ge, while it is only 0.35 $z_1$ in Si. Therefore, we 
obtain $3\beta/ \alpha = -1/3$ for Ge 
and $3\beta/ \alpha = -0.23$ for Si.

\begin{figure}[htbp]

\hspace*{+0.00cm} \epsfxsize=8cm \epsfbox{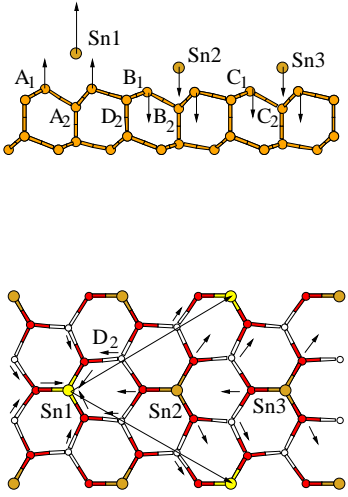}

\caption{ 
Ball-and-stick model of the Sn/Ge(111)--(3$\times$3) reconstruction.
The arrows show the direction of the atomic displacements (up/downwards on top, in-plane
on bottom figure) w.r.t. the $\sqrt{3}\times\sqrt{3}$--geometry. 
Notice that the Ge atoms  move in-plane towards the position of the upper ($Sn_1$) 
and away of the two other Sn atoms. The atoms D$_2$ connect the Ge atoms which are 
nearest neighbours of the  different Sn atoms.
}

\label{fig:r3to3x3}
\end{figure}

The phonon dispersion relation of this mode is easily found, using
eqn.~\ref{eqn:force}, to be:
\begin{equation}
M \omega^2 = (\alpha + 6 \beta) - \beta \sum_{j} cos (\bar{k} \cdot \bar{R}_{j})
\label{eqn:dispersion}
\end{equation}
where $\bar{k}$ is the momentum parallel to the surface, $\bar{R}_{j}$
the coordinates of the six first nearest neighbours, and $M$ the
tetrahedron mass.
In this approach,  the atoms outside the tetrahedra are treated
{\em adiabatically}, {\it i.e.}
we neglect their mass and assume them to
follow ``instantaneously" the displacement imposed by the  motion of the tetrahedra.
This means that the forces between, say, the Ge-D$_2$ atoms (see fig.
\ref{fig:r3to3x3}) and the atoms in the tetrahedra
are automatically included in the force constants of eqn.~\ref{eqn:force} and 
in the phonon dispersion relation of eqn.~\ref{eqn:dispersion}.  
We have estimated the error introduced by this adiabatic approximation in 
the phonon frequencies by comparing the 
kinetic energies of the atoms whose inertia is neglected in  our approach
with that of the atoms in the tetrahedra. This yields
errors of 7\% and 16 \% for the Si and Ge cases respectively. 
These values set the order of magnitude of the accuracy that our model 
presents for the phonon frequencies.

\begin{figure}[htbp]

\hspace*{+0.00cm} \epsfxsize=8cm \epsfbox{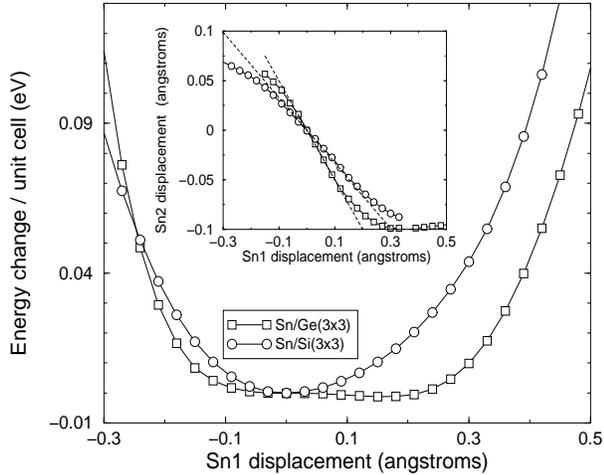}

\caption{ 
Total energy as a function of the $Sn_1$ for the Sn/Ge(111) and
Sn/Si(111) surfaces.
The energy of the $\sqrt{3}\times\sqrt{3}$ structure
is taken as the reference.
Inset: 
Displacements (in \AA) of the $Sn_2$, $Sn_3$ atoms as a function of
the $Sn_1$ displacement.
All the displacements are referred to the $\sqrt{3}\times\sqrt{3}$ structure.
}

\label{fig:energy_vs_Sn1}
\end{figure}

\begin{figure}[htbp]

\hspace*{-0.25cm} \epsfxsize=8cm \epsfbox{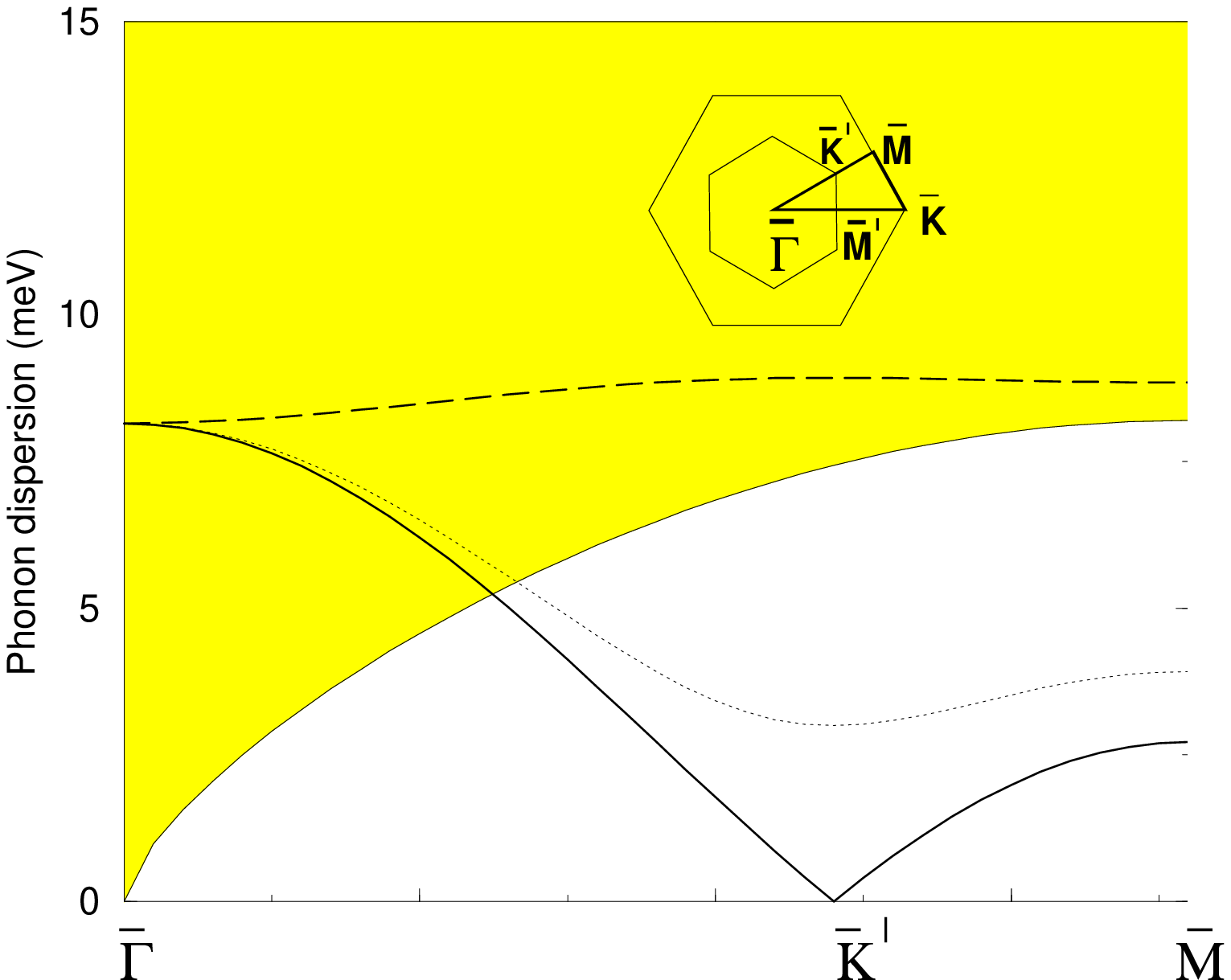}

\hspace*{-0.25cm} \epsfxsize=8cm \epsfbox{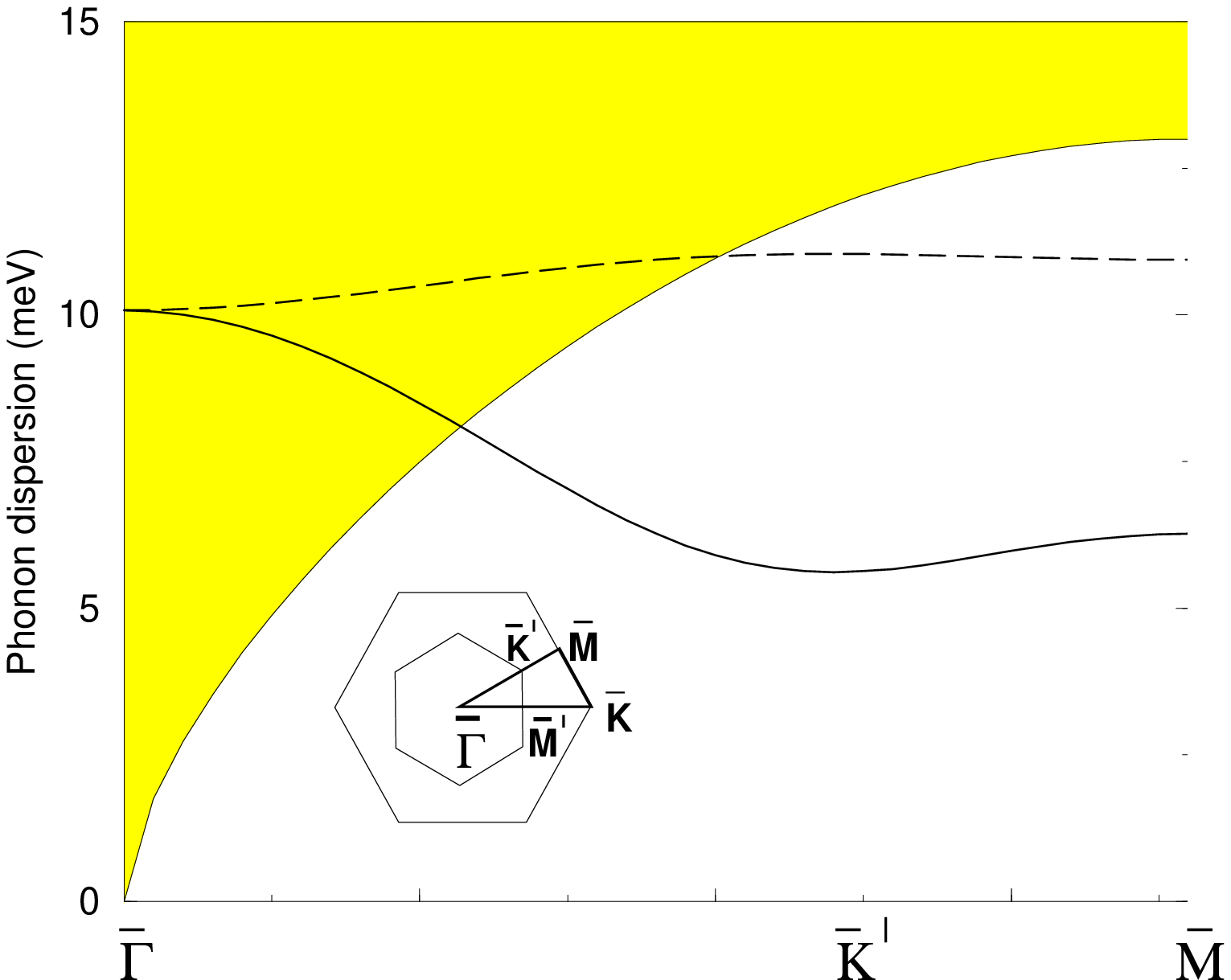}

\caption{ 
Phonon dispersion curves for the Sn/Ge(111) (full line, top) and 
Sn/Si(111)-$\sqrt{3}\times\sqrt{3}$
(full line, bottom) surfaces. Dashed lines correspond to the H-saturated
Sn/Ge(Si)(111)-$\sqrt{3}\times\sqrt{3}$ surfaces. 
Shaded areas represent the projection of the corresponding phonon bulk band
structure.
The phonon branches are plotted along the $\bar{\Gamma}\bar{{\rm M}}$
direction in the first Brillouin Zone (BZ) of the ideal (111) surface (see inset). 
The inner hexagon
corresponds to the BZ of the $\sqrt{3}\times\sqrt{3}$ surface. 
$\bar{K}^{\prime}$ defines
the new 3$\times$3 periodicity  associated with the soft phonon in Ge.
At room T, the frequency of the soft mode should be renormalized to a 
non-zero value, as shown schematically by the dotted line.
}

\label{fig:phonons}
\end{figure}

Figures~\ref{fig:phonons}a and \ref{fig:phonons}b show the phonon
dispersion curves
for Ge and Si, compared with the projected phonon bulk band structure. Notice
that we find a zero-frequency mode at the $\bar{K}^{'}$ point for Ge.
The corresponding $\bar{k}$--vector defines the new 3$\times$3 periodicity
associated with this soft mode. For Si, the phonon dispersion relation
shows only a minimum at the $\bar{K}^{'}$ point, with a phonon energy of 5.5
meV. These results are in agreement with our previous comments on the
stability of the Sn/Si(111)-$\sqrt{3}\times\sqrt{3}$ structure, as shown in
Figure~\ref{fig:energy_vs_Sn1}.

The phonon branches shown in Fig.~\ref{fig:phonons} provide an explanation for
the main similarities and differences between the Si and Ge cases. At low T, 
while for Ge we can expect a phase transition to the 3$\times$3 structure, 
there is no stable 3$\times$3 phase for Si.  
On the other hand, at high T (say, for $k_{B}T >> 6$ meV, room T is a high T
limit) we can expect Ge and Si to be alike. This can be understood by
considering the energy curves in Fig.~\ref{fig:energy_vs_Sn1}: 
at room T the Sn atoms vibrate with a
large amplitude in both surfaces (with displacements between -0.2 and 0.3
\AA\ for Ge and -0.2 and 0.2 \AA\ for Si).
In this vibration, Sn atoms are correlated with their
Sn--nearest--neighbours.
This is true even in
the Si case, because although the 3$\times$3 structure is not an stable minimum,
due to the fact that the
vibration associated with the 3$\times$3 structure is a minimum of the phonon
dispersion, the atomic motion can be expected to be dominated by this mode.
For the Sn/Ge(111)--$\sqrt{3}\times\sqrt{3}$ phase,
anharmonic effects renormalize
the frequency of the soft-phonon mode
to a non-zero value\cite{vanderbilt}, as shown
schematically in Figure~3.

The atoms involved in this vibration spend most of 
the time around the turning points of their classical trajectories,
where the atom velocity goes to zero.
At these turning points, the electron occupation of the Sn dangling bonds 
is different: atoms in ``up" positions have a fully occupied dangling bond,
while those at ``down" positions present only a partial occupation of 
their dangling bonds\cite{ortega}. 
This explains the double peak observed in the XPS Sn
$4d$ core level spectra of both systems,
as originally proposed in the dynamical fluctuations
model\cite{avila} to explain
the Sn/Ge(111) $3\times 3 \rightarrow \sqrt{3}\times\sqrt{3}$ transition.      
In this model, Sn atoms display correlated up/down vibrations which
present, at RT, large amplitudes, but keep memory of the underlying 
3$\times$3 phase.
The soft phonon shown in fig.~\ref{fig:phonons} is responsible for 
this atomic motion and provides the physical mechanism for the dynamical 
fluctuation model.

We have also addresed the problem of which driving force is softening
the surface mode associated with the 3$\times$3 structure.
As Si and Ge are isovalent atoms and the electronic structure of both interfaces
is very similar, one might expect the difference in the
relative stability of the two structures to be related to the different elastic
properties of the two materials.
In order to single out the elastic contribution from the
one associated with the Sn--dangling--bond rehybridization \cite{tosatti} and
corresponding surface--band splitting\cite{ortega},
we have studied the surface phonons of these surfaces
with hydrogen atoms saturating the Sn--dangling--bonds. 
As shown in Figs.~\ref{fig:phonons}a and \ref{fig:phonons}b,
the dispersion curves for the H--saturated 
surfaces are rather different
from the Sn--surfaces; in particular they display no
``softening" of the mode at $\bar{K}^{'}$.
This different behaviour is related to the
Sn--dangling--bonds rehybridization.

In order to quantify these effects, we have considered the relaxed
Sn--surfaces discussed above (see Fig. 2)
for two different $Sn_1$ displacements, namely, 0.0 \AA
(the $\sqrt{3}\times\sqrt{3}$ geometry) and 0.18 \AA
(the ``3$\times$3" geometry).
The elastic contribution is calculated, in each case,
as the difference in energy between the two geometries {\it with H--atoms
saturating the Sn--dangling--bonds}.
Substracting this elastic contribution from the total energy difference for the
Sn--surfaces, the energy contribution due to the Sn--dangling--bonds
rehybridization is obtained.
These calculations provide  the following results: $E_{elastic}
(Ge) = 156$
meV and $E_{electronic} (Ge) = -158$ meV; $E_{elastic} (Si) = 131$ meV 
and $E_{electronic} (Si) = -112$ meV \cite{elastic}.     
Therefore,
the main difference between the Sn/Ge and Sn/Si cases is
the different hybridization between Sn and the
surface atoms.
A comparison of the surface bands for both cases confirms this result:
the Ge case presents in the 
3$\times$3 geometry a larger
band splitting, 152 meV as compared with 93 meV for Si.
This effect can be 
related to the different sizes of Si and Ge \cite{tosatti}.
When Sn is closer in size
to the semiconductor atom the surface rehybridization
is stronger and the surface
gains more energy by means of the surface band splitting caused by the
3$\times$3 distortion.


In conclusion, we have shown via state-of-the-art DFT calculations that 
the reversible $\sqrt{3}\times\sqrt{3} \leftrightarrow 3\times3$
phase transition in Sn/Ge(111) is associated with a surface soft mode.
The $3\times3$ structure is not stable in the Sn/Si(111) case, although
it is a minimum in the phonon dispersion, and consequently this
surface should not display the 
$\sqrt{3}\times\sqrt{3} \leftrightarrow 3\times3$ transition.
Comparing the Sn/Ge and Sn/Si cases, and considering the corresponding  
H-covered surfaces, we have found that the driving force underlying the 
$\sqrt{3}\times\sqrt{3} \leftrightarrow 3\times3$
phase transition lies in the electronic energy gain due to 
the Sn dangling bond rehybridization.

This work has been partly funded by the spanish CICYT under contract
No.PB-97-0028. We acknowledge helpful discussions with V.R. Velasco and
J.M. P\'erez-Mato. 


\begin{table}

\caption{Atomic displacements (in \AA) associated with the
Sn/Ge(111)--(3$\times$3) surface measured w.r.t. the $\sqrt{3}\times
\sqrt{3}$--geometry.}

\begin{tabular}{ddddddd}
\ & $Sn_1$ & $Sn_2$ & $Sn_3$ & A$_1$ & B$_1$ & C$_1$  \\
\hline
$\Delta z$ & 0.18 & -0.08 & -0.08 & 0.08 & -0.04 & -0.03 \\
$\Delta \perp $ & 0.00 & 0.00 & 0.00 & -0.04 & 0.02 & 0.02 \\
\end{tabular}                                                                   

\begin{tabular}{ddddd}
\ & A$_2$ & B$_2$ & C$_2$ &  D$_2$ \\
\hline
$\Delta z$ & 0.02 & 0.00 & 0.00 &  0.00 \\
$\Delta \perp $ & 0.00 & 0.00 & 0.00 &  0.04 \\
\end{tabular}                                                                   

\label{tab:r3to3x3}
\end{table}

\begin{references}
\bibitem{elastic} 
$E_{elastic} (Si)$ is smaller than $E_{elastic}(Ge)$ due to the smaller
displacements of the Si-atoms, for the same $Sn_1$ displacement.
\end{references}
\end{document}